\documentclass[aps,prb,reprint,twocolumn,superscriptaddress]{revtex4-2}
\usepackage{graphicx}
\usepackage{amsmath}
\usepackage{amsfonts}
\usepackage{color}
\usepackage{amssymb}%
\usepackage{siunitx}
\usepackage{blindtext}
\usepackage{xspace}
\usepackage{siunitx} %for Units in APS journals like PRL and PRB (https://tex.stackexchange.com/questions/9647/using-siunitx-when-submitting-to-aps)
\usepackage{miller}

\usepackage[version=3]{mhchem}
\usepackage{natbib}
\usepackage[]{hyperref}

\usepackage{bm} 
\usepackage{orcidlink}

\renewcommand{\vec}[1]{{\boldsymbol #1}}
\newcommand{\tbmno}{TbMnO$_3$ }

\newcommand{\mnwo}{MnWO$_4$ }
\newcommand{\mnwokomma}{MnWO$_4$}

\newcommand{\nh}{$\left(\text{NH}_4\right)$$_2$$\left[\text{FeCl}_5(\text{H}_2\text{O})\right]$ }
\newcommand{\Ah}{$A$$_2$$\left[\text{FeCl}_5(\text{H}_2\text{O})\right]$ }

\newcommand{\nhkomma}{$\left(\text{NH}_4\right)$$_2$$\left[\text{FeCl}_5(\text{H}_2\text{O})\right]$}
\newcommand{\nd}{$\left(\text{ND}_4\right)$$_2$$\left[\text{FeCl}_5(\text{D}_2\text{O})\right]$ }
\newcommand{\ndkomma}{$\left(\text{ND}_4\right)$$_2$$\left[\text{FeCl}_5(\text{D}_2\text{O})\right]$}

\newcommand{\nafegeo}{NaFeGe$_2$O$_6$ }
\newcommand{\nafegeokomma}{NaFeGe$_2$O$_6$}

\newcommand{\erykomma}{erythrosiderite}

\begin{document}

%\preprint{}

\title{Multiferroic domain relaxation in \nh}

\author{S. Biesenkamp}
\affiliation{$I\hspace{-.1em}I$. Physikalisches Institut,
Universit\"at zu K\"oln, Z\"ulpicher Straße 77, D-50937 K\"oln,
Germany}

\author{K. Schmalzl}
\affiliation{Juelich Centre for Neutron Science JCNS, Forschungszentrum Juelich GmbH, Outstation at ILL, 38042 Grenoble, France}

\author{P. Becker}
\affiliation{Institut  f\"ur  Geologie  und  Mineralogie, Abteilung  Kristallographie, Universit\"at zu K\"oln, Z\"ulpicher Straße  49b,  50674  K\"oln,  Germany}

\author{L. Bohat\'{y}}
\affiliation{Institut  f\"ur  Geologie  und  Mineralogie, Abteilung  Kristallographie, Universit\"at zu K\"oln, Z\"ulpicher Straße  49b,  50674  K\"oln,  Germany}

\author{M. Braden\,\orcidlink{0000-0002-9284-6585}}\email[e-mail: ]{braden@ph2.uni-koeln.de}
\affiliation{$I\hspace{-.1em}I$. Physikalisches Institut,
Universit\"at zu K\"oln, Z\"ulpicher Straße 77, D-50937 K\"oln,
Germany}



%\thanks{}


\date{\today}

\begin{abstract}
The molecular compound $\left(\text{NH}_4\right)$$_2$$\left[\text{FeCl}_5(\text{H}_2\text{O})\right]$ is a type-II multiferroic material, in which incommensurate cycloidal order directly induces ferroelectric polarization.
The multiferroic domain kinetics in $\left(\text{NH}_4\right)$$_2$$\left[\text{FeCl}_5(\text{H}_2\text{O})\right]$ were studied by time-resolved neutron diffraction experiments utilizing neutron polarization analysis.
The temperature and electric-field dependent multiferroic relaxation obeys the simple combined Arrhenius-Merz law, which was reported to describe domain kinetics in the prototype multiferroics \tbmno and \nafegeokomma{.} However, the characteristic time scale of the multiferroic relaxation is considerably larger than those in TbMnO$_3$ or NaFeGe$_2$O$_6$.
Temperature-dependent diffraction on \nh reveals the emergence of higher-order and commensurate magnetic contributions upon cooling in the multiferroic phase in zero field.
The good agreement with studies of higher-harmonic contributions in the deuterated material indicates that the isotopes only posses a minor impact on the magnetic ordering.
But in contrast to similar observations in multiferroic \mnwokomma{,} this anharmonic modification of magnetic ordering does not depin multiferroic domain walls or alter the temperature dependence of the multiferroic relaxation.

\end{abstract}

%\keywords{}

\maketitle


\section{\label{sec:level1}Introduction}
%The leap into the digital age entailed permanent progression and fine-tuning of up-to-date methods and %devices that are used by information and communication technology.

The control of complex magnetic and electric ordering parameters by external electric or magnetic fields stimulates the research on multiferroic materials \cite{Khomskii2009,Spaldin2019,Scott2007a}. Especially type-II multiferroics attract much interest, as improper ferroelectric ordering is not only coexisting with magnetic ordering but is also induced by it \cite{Khomskii2009}. This inherently entails strong magneto-electric coupling and an application potential in low-power memory devices \cite{Scott2007a}. The inverse Dzyaloshinskii-Moriya interaction (DMI) \cite{Dzyaloshinsky1958,Moriya1956} drives the multiferroic state in most type-II multiferroics. The canting of neighboring spins in spirals induces a shift of non-magnetic ions in between and thus a ferroelectric polarization, whose sign depends on the handedness of the spiral \cite{Mostovoy2006}. An external electric field can thus control the sign of the vector chirality of the spiral spin structure.

Recent studies of multiferroic domain inversion revealed that the relaxation in \tbmno and \nafegeo follows a simple combined Arrhenius-Merz law suggesting thermally activated domain-wall motion \cite{stein2020,biesenkamp2021b}. 
In the general Arrhenius activation, the dynamics of a process is described by $\exp(-\frac{E_A}{k_BT})$ or by a relaxation time following $\tau\propto \exp(\frac{E_A}{k_BT})$ with $E_A$ the activation energy.
In a ferroelectric material the electric field can control and invert the domains, and in many compounds the dependence of this relaxation on the electric field follows the Merz law at constant temperature, $\tau\propto \exp(\frac{F_A}{E})$ where $F_A$ is an activation field \cite{scott2000}.
However, in proper ferroelectric switching, the nucleation at the surface, their subsequent forward growth along the polarization direction and finally the  sideways growth of needle shaped domains act on the same time scale \cite{scott2000} rendering the temperature dependence difficult to describe.
In contrast, the multiferroic domain inversion is mostly dominated by the sideways growth of some few large domains \cite{Meier2009,Hoffmann2011a,Matsubara2015,stein2020}.
In \tbmno and in \nafegeo the combination of both relations well describes the multiferroic relaxation over many decades in time.
The combined Arrhenius-Merz relation is expressed as:

\begin{align}
  \label{eq:arrhenius-merz}
\tau\left(E,T\right)=\tau^*\text{exp}\left(\frac{A_0T_r}{ET}\right)\:\text{with}\:T_r=\frac{T_\text{MF}-T}{T_\text{MF}}.
\end{align}

In this relation there are only two independent parameters: an activation constant $A_0$ and a critical relaxation time $\tau^*$ that is reached
by either increasing the electric field to infinity or by approaching the multiferroic transition, $T_r=0$. 
The observation of such a simple relation over many time decades indicates a single slow process to dominate the multiferroic relaxation, i.e. the growth of a few large domains by domain wall motion.
Furthermore, the critical relaxation time $\tau^*$ must correspond to some intrinsic entity that does not change in the multiferroic phase. 
It is tempting to identify $\tau^*$ with the inverse of the spin-wave velocity, but the effective distance the domain growth must cover is
neither known nor controlled.

There is at least one exception to this relation. The domain inversion in multiferroic \mnwo is peculiar as it speeds up 
upon cooling towards the lower boundary of the multiferroic phase where the incommensurate order transforms into a commensurate one \cite{Baum2014,Niermann_2014,Hoffmann2011a}.
Magnetic correlations in \mnwo start to resemble the commensurate spin up-up-down-down ($uudd$) arrangements already in its incommensurate multiferroic phase, as tiny commensurate fragments occur.
Therefore, the interference of commensurate and incommensurate ordering can depin multiferroic domains and thus cause the speeding up of multiferroic domain inversion near the lower first-order magnetic transition \cite{biesenkamp2020}.
The different relaxation behavior motivates the investigation of multiferroic domain inversion in other materials, in particular in a molecular
compound, which promises a softer nuclear lattice.

The \erykomma{s} \Ah with $A$ being an ammonium or an alkali-metal ion have attracted much interest as some members exhibit magnetoelectric and even multiferroic behavior \cite{Ackermann_2013,Ackermann_2014}. Structurally, \erykomma{s} are closely related to each other.
The magnetic Fe$^{3+}$ ion is surrounded by 5 Cl$^-$ ions and the O of the water molecule H$_2$O.
These FeCl$_5$H$_2$O octahedra are not directly connected (no shared ligands) but separated by the $A$ ions. 
The hydrogen of the H$_2$O ligands, however, provides bonding between O and Cl ions of neighboring octahedra. 
Considering these H-bridge bonds the FeCl$_5$H$_2$O octahedra form zigzag chains. 
However, the alignment of these zigzag chains differs in the material class between $A$\,=\,Cs and the others ($A=\text{NH}_4\text{, Rb, K}$) \cite{lindqvist1947,Lackova2013,Figgis1978,Connor1979,Greedan1980,Schultz1995}.
The latter compounds crystallize in space group $Pnma$ with chains along $b$, see Fig. 1, and
the H-bonds are fully determined by the general arrangement. In contrast for $A$\,=\,Cs, there is
a structural phase transition, at which these bonds order \cite{Frohlich2018}.
Most \erykomma{s} were reported to develop antiferromagnetic and collinear magnetic structures at low temperatures \cite{Gabas_1995,Campo2008,luzon2008}, whereby the easy axis points along the $a$ direction. However, this does not hold for the ammonium compound, for which an $ac$ easy plane with $XY$ anisotropy was observed \cite{Velamazan2015,Tian2016}.

Orthorhombic \nh (space group $Pnma$) exhibits the unit cell dimensions $a=\SI{13.706(2)}{\angstrom}$, $b=\SI{9.924(1)}{\angstrom}$ and $c=\SI{7.024(1)}{\angstrom}$ \cite{McElearney1978,Ackermann_2013} at room temperature. 
At $T_\text{S}\approx\SI{79}{\kelvin}$ a structural order-disorder transition of the tetrahedral $\text{(NH)}_4^+$ groups leads to the loss of $n$ and $m$ symmetries and thus to a monoclinic distortion of the crystal lattice \cite{bruening2020,Velamazan2015}. Below $T_\text{S}$ the system can be described by space group $P2_1/a$ (nonstandard setting) and no further symmetry reductions are reported at low temperature \cite{bruening2020}.
\begin{figure}
 \includegraphics[width=\columnwidth]{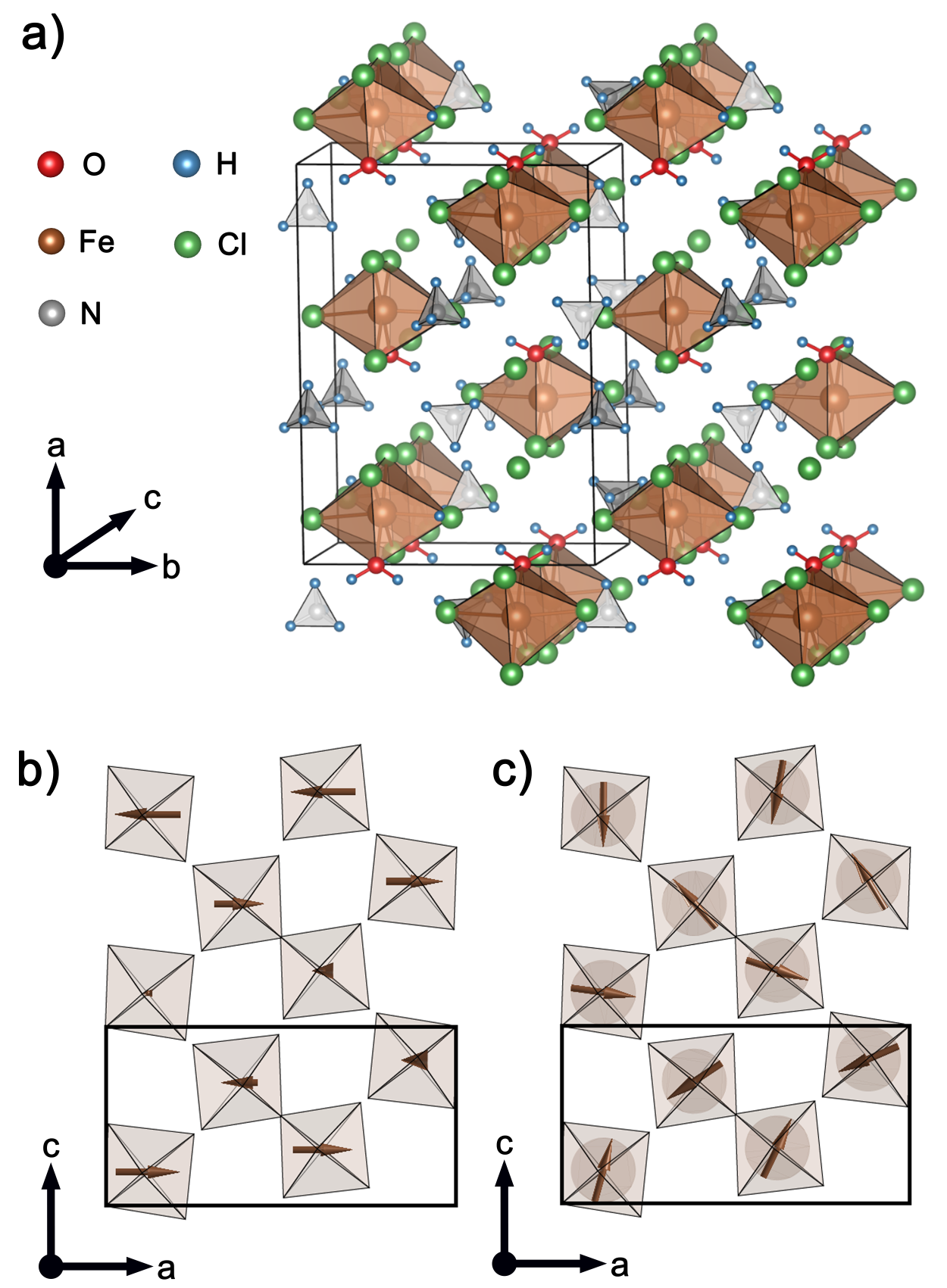}
  \caption{\label{fig:structure}Panel a) displays the crystal structure of \nh with ordered $\text{(NH)}_4^+$ tetrahedra, whereby the respective structural data were taken from the low-temperature refinement in Ref. \onlinecite{bruening2020}. For the visualization of the crystallographic parts the software VESTA3 \cite{Momma2011} was utilized.  In b) and c), the magnetic structure of both phases, the intermediate SDW phase and the low-temperature spiral multiferroic phase are sketched within the $ac$ plane. The magnetic structure information was taken from Ref. \onlinecite{Velamazan2015} and Ref. \onlinecite{Tian2016}.}
  \label{structure}
 \end{figure}
At $T_\text{N}\approx\SI{7.3}{\kelvin}$ long-range magnetic order in the form of an incommensurate spin density wave (SDW) with moments along the $a$ direction sets in \cite{Tian2016,Ackermann_2013,Velamazan2015}. The propagation vector $\vec{k}=(0\:\:0\:\:0.23)$ displays a slight temperature dependence but it does not reach a commensurate value \cite{Tian2016}. 
Below the multiferroic transition at $T_\text{MF}$$\approx\SI{6.8}{\kelvin}$ a spiral spin structure forms with moments rotating in the $ac$ plane \cite{Velamazan2015,Tian2016} and ferroelectric polarization of $\approx\SI{3}{\micro\coulomb\per\square\metre}$ emerges \cite{Ackermann_2013}. 
The substitution of the protons by deuterium slightly enhances the structural transition by about one Kelvin but has no measurable impact on the two magnetic transitions \cite{bruening2020}.
Note, that the ferroelectric polarization arising from the magnetic ordering was only determined for the proton material \cite{Ackermann_2013}.
The inverse DMI well explains multiferroic coupling in \nhkomma{.} However, Tian \textit{et al.} reported that the groundstate is not homogeneous but consists of two soft phases in form of incommensurate and commensurate ordering \cite{Tian2018}. 
In zero magnetic field, incommensurate ordering predominates but at a magnetic field along $a$ of $\approx\SI{2.5}{\tesla}$ the structure transforms into a distorted commensurate cycloid \cite{Tian2018,Velamazan2018}, and at $\approx\SI{5}{\tesla}$ a spin-flop transition induces a commensurate canted spin structure with moments lying mainly within the $bc$ plane \cite{Velamazan2018,Ackermann_2013}.
The ferroelectric polarization at zero field has its main component along the $a$ direction with a ten times smaller component along $b$ \cite{Ackermann_2013}.
At the first transition in magnetic fields the polarization tilts slightly away from the $a$ direction, and above the spin-flop transition the ferroelectric polarization points along the $c$ direction. 
This orientation of ferroelectric polarization disagrees with the DMI mechanism but can be described by the spin dependent $p$-$d$ hybridization model \cite{Velamazan2018,Ackermann_2013}. 
The interfering of incommensurate and commensurate ordering in the multiferroic phase as well as the molecular character promise different relaxation processes and motivate the investigation of the domain kinetics in \nhkomma{.}

In the following sections we first introduce the experimental methods before presenting time-resolved polarized neutron-diffraction studies of multiferroic domain inversion in \nhkomma{.} The temperature and electric field dependence of relaxation times follows again the combined Arrhenius-Merz relation given in equation  \ref{eq:arrhenius-merz} \cite{stein2020}. In addition, we discuss the temperature dependence of the commensurate correlations and that of higher harmonic reflections and their possible impact on the relaxation behavior.

\section{Experimental methods}\label{sec:expmethods}

For the investigation of the multiferroic domain dynamics and for sensing the handedness of the spiral spin structure, neutron diffraction with polarization analysis was utilized. As the molecular compound \nh contains hydrogen atoms, which cause a strong incoherent background, previous magnetic neutron-scattering experiments were performed only on  deuterated samples \cite{Velamazan2015,Tian2016,Tian2018,Velamazan2017,Velamazan2018,bruening2020}. Polarized neutron diffraction does not significantly profit from the reduced background signal, and as the handling of deuterated samples is complicated we prepared two non-deuterated samples (SI and SII), with dimensions of about $\SI{3.8}{\milli\metre}\times\SI{4.8}{\milli\metre}\times\SI{1.03}{\milli\metre}$  for SI and $\SI{5.2}{\milli\metre}\times\SI{5.8}{\milli\metre}\times\SI{3.1}{\milli\metre}$ for SII. 
%This also enables  an investigation of isotope effects on the magnetic structure. 
An elaborate description of the sample growth can be found in Ref. \onlinecite{Ackermann_2013}.  Both crystals were clamped between alumiunium plates that are tightened together by insulating polytetrafluorethylen (PTFE) screws. The plate normal was oriented parallel $a$ in order to apply the electric field along the predominant component of the ferroelectric polarization. Both plates were mounted on separate sample holders in the scattering geometry (0 1 0)/(0 0 1) and connected to the high-voltage setup. This setup contains two \textsc{Iseg} modules (\textsc{BPP4W} and \textsc{BPN4W}), which provide a maximum output voltage of plus and minus $\SI{4}{\kilo\volt}$, respectively, and a MOSFET array
(\textsc{Behlke HTS-111}) that is capable to switch periodically between both polarities within less than $\SI{50}{\micro\second}$ (for a detailed description of this setup see Ref. \onlinecite{stein2020}) generating a nearly rectangular time shape of the electric-field profile. 
The time-resolved measurements are realized with a multichannel
data collector (\textsc{Mesytec MCPD-8}), which records neutrons in event mode with a timestamp. 
After the data recording single neutron events are distributed to the discrete time dependency $I( [t_i,t_i+\Delta t] )$ with a time binning $\Delta t$ that can still be adapted.
Very fast relaxation processes can be resolved, because the statistics only depend on the number of switching periods and hence can be increased at any order. 
For measuring quasistatic hysteresis loops, while driving the applied electric field between its extrema, a \textsc{Fug HCP 14-3500} high-voltage generator was used.

The time-resolved investigations of the multiferroic domain inversion were carried out at the triple-axis spectrometer IN12 at the Institute Laue-Langevin (ILL) \cite{Schmalzl2016}. A pyrolytic graphite (PG(002)) monochromator supplied $\lambda=\SI{3.14}{\angstrom}$, while a cavity polarizer provided a highly polarized neutron beam. The sample was inserted in a standard orange cryostat and a Helmholtz-coil setup defined the guide-field direction at the sample position. Two spin flippers that are  positioned before and behind the sample as well as a Heusler (111) analyzer enabled longitudinal polarization analysis. The flipping ratio (FR) was measured on the (0 2 0) nuclear reflection and amounts to $\text{FR}\approx 20$.

For measuring the temperature dependence of the higher harmonics, we carried out an experiment at the IN3 triple-axis spectrometer at the ILL. A PG (002) monochromator provided an unpolarized neutron beam with $\lambda=\SI{2.36}{\angstrom}$ and the background signal was significantly reduced by a PG (002) analyzer. With the standard orange cryostat, the temperature dependence of higher harmonics was measured down to $T=\SI{1.5}{\kelvin}$.

\begin{figure}
 \includegraphics[width=\columnwidth]{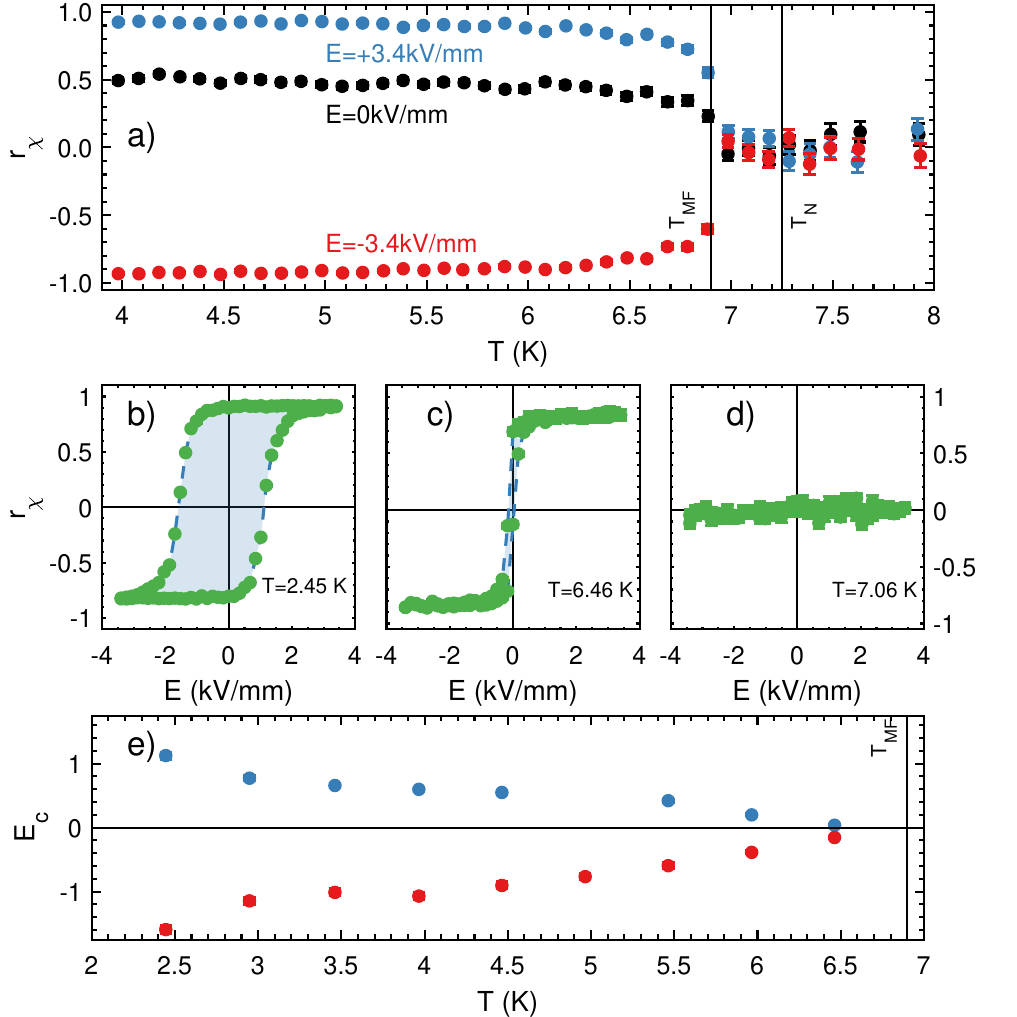}
  \caption{\label{fig:poling}The temperature dependence of the chiral ratio for different poling fields is displayed in a). Panels b), c) and d) show three exemplary  hysteresis loops at $T=\SI{2.45}{\kelvin}$, $T=\SI{6.46}{\kelvin}$ and $T=\SI{7.06}{\kelvin}$. The coercive fields were obtained by fitting both slopes with a hyperbolic tangent and the respective temperature dependence is shown in e). All measurements were recorded for the $\vec{Q}=(0\:\:1\:\:0.23)$ reflection.}
  \label{poling}
 \end{figure}

 \begin{figure}
 \includegraphics[width=\columnwidth]{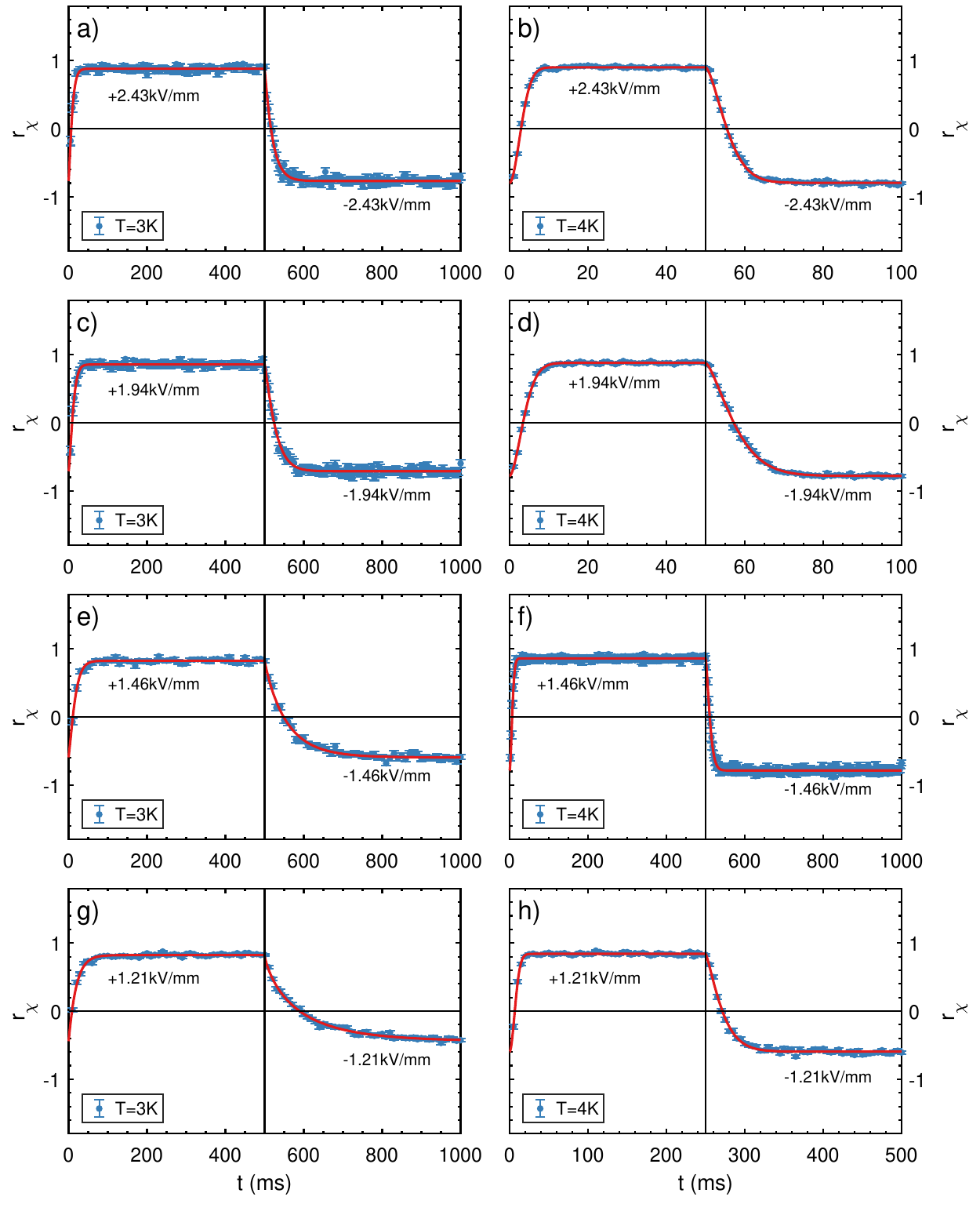}
  \caption{\label{fig:switching}Multiferroic switching curves are plotted in panels a)-h) for different temperatures and electric field amplitudes. These measurements were performed on sample SI and by recording both SF channels $I_{x\bar{x}}$ and $I_{\bar{x}x}$ for the magnetic reflection $\vec{Q}=(0\:\:1\:\:0.23)$.  }
  \label{switching}
 \end{figure}

\section{Multiferroic domain control}\label{sec:quasistatic}

For the temperature and electric field dependent investigation of the multiferroic domain dynamics, sample SI was measured at the IN12 triple-axis spectrometer utilizing longitudinal polarization analysis.
Unpolarized neutron diffraction senses the Fourier transform of the magnetization perpendicular to the scattering vector  $\vec{M}_\perp$
yielding a magnetic intensity contribution proportional to $|\vec{M}_\perp|^2$. Without neutron polarization, there is no interference between nuclear
and magnetic structures, but,
for a polarized neutron beam, additional terms need to be considered \cite{Brown2006}. Since we are studying purely magnetic Bragg peaks, there is no nuclear
magnetic interference, but the chiral term, $i\left(\vec{M}_\perp\times\vec{M}_\perp^*\right)$,
causes magnetic cross sections that depend on the sign of the neutron polarization both before and after the scattering process.
We use the common coordinate system in polarized neutron diffraction, where ${\bf x}$ is along the scattering vector ${\bf Q}$, ${\bf y}$ perpendicular
to ${\bf Q}$ in the scattering plane and ${\bf z=x\times y}$. 
The chiral term can be studied by measuring both spin-flip (SF) channels $I_{x\bar{x}}$ and $I_{\bar{x}x}$ yielding the chiral ratio $r_\chi=\left(I_{x\bar{x}}-I_{\bar{x}x}\right)/\left(I_{x\bar{x}}+I_{\bar{x}x}\right)=\pm i\left(\vec{M}_\perp\times\vec{M}_\perp^*\right)_x/|\vec{M}_\perp|^2$.
The sign of the chiral contribution senses the sign of the vector chirality and thus the handedness of the spin spiral.
In an ideal scattering geometry with ${\bf Q}$ perpendicular to the envelope of the chiral magnetic structure and assuming a circular rotation of
magnetic moments, a chiral ratio of $\pm 1$ is expected for a monodomain state.
If the scattering vector is not perfectly perpendicular to the rotation plane of spins or if the spiral is elliptically distorted, the maximum value of $r_\chi$ is reduced, but rather large deformations are required to yield a significant deviation of the expectation $r_\chi=1$ for a monodomain state.

In the first part of the experiment, the chiral ratio was measured at $\vec{Q}=(0\:\:1\:\:0.23)$ as a function of temperature and for different applied electric fields. For each run, the sample was heated above the N\'eel temperature and subsequently cooled with applied electric field, while recording both SF channels $I_{x\bar{x}}$ and $I_{\bar{x}x}$. In Fig. \ref{fig:poling} a) the temperature dependence of $r_\chi$ is shown for $E=\pm\SI{3.4}{\kilo\volt\per\milli\meter}$ as well as for zero applied field. The values clearly state that the sign of the vector chirality can be poled in opposite directions depending on the applied field direction and even close to the multiferroic transition temperature the system can be poled to a monodomain state. Moreover, it can be seen that in zero field a finite value for $r_\chi$ develops, which indicates a preferred state of the multiferroic domain distribution for this sample.
In consequence, multiferroic domains at zero electric field should already be rather large.

The temperature dependence of the chiral ratio shown in Fig. \ref{fig:poling} a) nicely agrees with that of the ferroelectric polarization \cite{Ackermann_2013}, as it is expected for the inverse Dzyaloshinski-Moriya coupling. Both sense the product of the two orthogonal spin components of the cycloidal structure.
Upon cooling, both quantities exhibit a small jump at $T_\text{MF}$ and then further increase in the multiferroic phase.

Hysteresis  loops were recorded by quasistatically driving the applied electric field between $E=\pm\SI{3.4}{\kilo\volt\per\milli\meter}$ for different temperatures. Three exemplary loops are shown in Fig. \ref{fig:poling} b)-d). The hysteresis loops confirm the invertibility of the spiral handedness in \nh by external electric fields, while the visible asymmetry of the coercive fields agrees with the results from the poling sequences. 
Figure \ref{fig:poling} e) presents the temperature dependence of the coercive fields $E_\text{c}$ obtained by fitting both slopes of the hysteresis loop by a hyperbolic tangent. 
The quasistatic measurements resemble those on a deuterated sample \cite{Velamazan2018}, which however
revealed smaller values of the chiral ratios. 
Upon cooling through the multiferroic transition with applied field we find $r_\chi$\,=\,0.920(6) and also
in the hysteresis a high value is reached at the favorite side, $r_\chi$\,=\,0.92(1). 
For a perfect circular envelope and a monodomain state one expects an ideal chiral ratio of 1 that is reduced by the finite flipping ratio to $r_\chi$=0.91 in
perfect agreement with the observations.
The alignment of multiferroic domains is thus perfect within a few percent.

\section{Relaxation behavior}\label{sec:switching}

The relaxation times of multiferroic domain inversion as a function of temperature and electric field  were determined by deploying the time-resolved setup (see Ref. \onlinecite{stein2020}) at the IN12 spectrometer. 
These measurements were performed on the same sample and on the same position in Q space as for the quasi-static measurements.
Figure \ref{fig:switching} displays exemplary switching curves that were recorded for different temperatures and electric field amplitudes. 
It can be seen that the inversion time for these switching curves cover more than one order of magnitude as a function of temperature and electric field. 
Furthermore, the observed preference of one domain population becomes evident by the asymmetry of the switching curve. 
In analogy to the analysis in Ref. \onlinecite{stein2020} and Ref. \onlinecite{biesenkamp2021b} the rising and falling parts of all switching curves were fitted by stretched exponential functions:

\begin{align}
  \label{eq:switching}
r_\chi(t)&=r_a-\left(r_a-r_b\right)\text{exp}\left(-\left(\frac{t}{\tau_a}\right)^{b_1}\right)\\
r_\chi(t)&=r_b-\left(r_b-r_a\right)\text{exp}\left(-\left(\frac{t-t_{1/2}}{\tau_b}\right)^{b_2}\right)\label{eq:switching_2}
\end{align}
  \begin{figure}
 \includegraphics[width=\columnwidth]{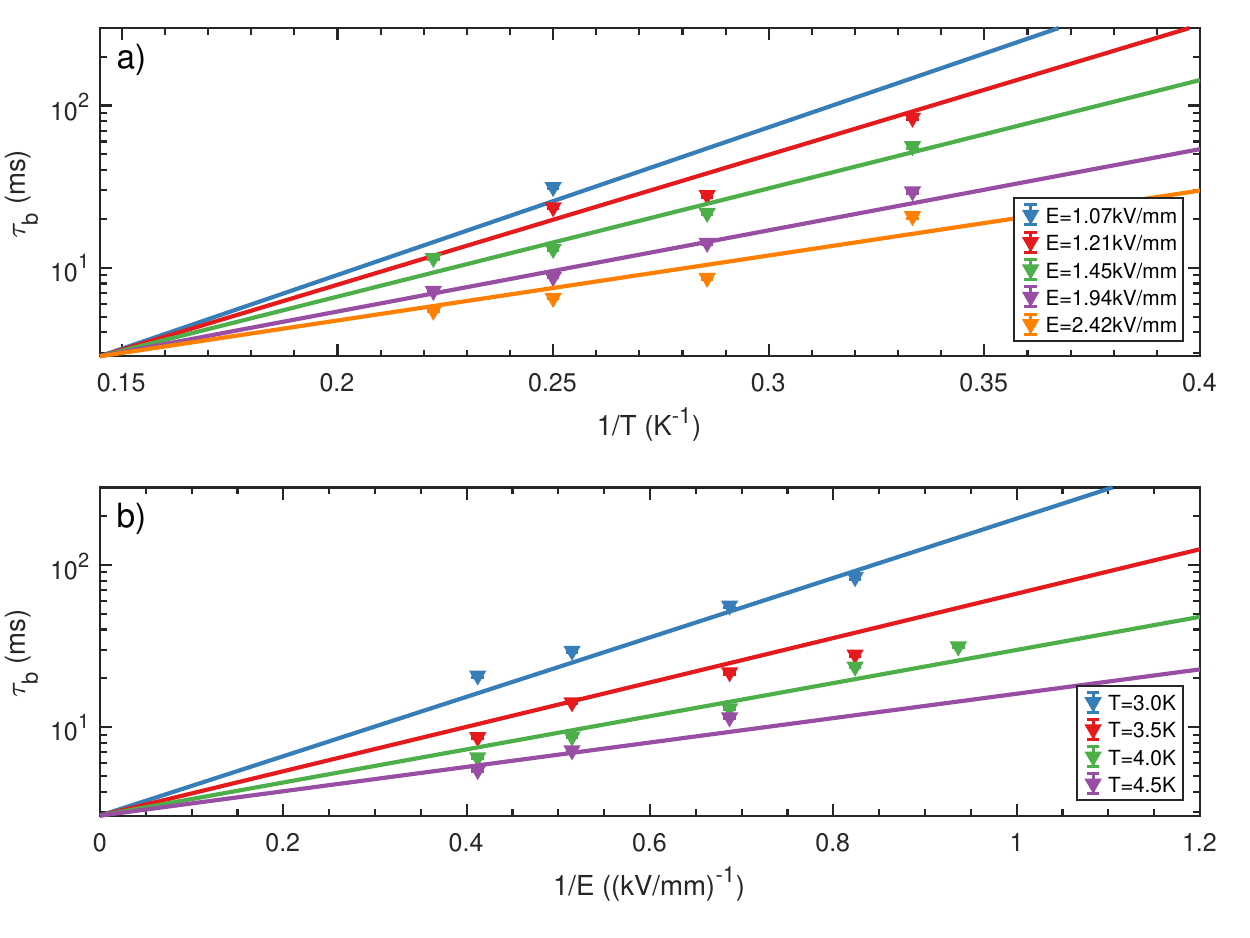}
  \caption{\label{fig:relaxation_times} Both panels a) and b) display the determined relaxation time on a logarithmic scale as a function of inverse temperature and inverse electric-field amplitude. Because of the observed asymmetry, the relaxation time is plotted only for one switching direction (here $\tau_b$). Solid lines refer to the fit result for respective temperatures and fields. }
  \label{relaxation}
 \end{figure}

These functions describe the relaxation process from state $r_b$ to state $r_a$ and vice versa with relaxation times $\tau_a$ and $\tau_b$, respectively.
The value $t_{1/2}$ denotes the switching time of the external field at half the period. 
Following the Ishibashi and Takagi theory \cite{Ishibashi_1971,Ishibashi_1990}, which relies on the Avrami model for phase change kinetics \cite{Avrami_1939,Avrami_1940,Avrami_1941}, both exponential factors $b_1$ and $b_2$ indicate the domain growth dimensionality for respective switching directions. 
Fitting all switching curves with equation \ref{eq:switching} and \ref{eq:switching_2} yield $b$ parameters in the range between 1 and 2, which indicates low dimensional domain growth. 
Furthermore, the b values tend to raise for increased field or temperature. 
This is expected as for higher temperatures or higher electric fields the assumption of only a few nuclei forming at the beginning transforms into a continuous growth of nuclei during the switching process. 
An equivalent behavior was observed for \tbmno and \nafegeo \cite{stein2020,biesenkamp2021b}.

   \begin{figure}
 \includegraphics[width=\columnwidth]{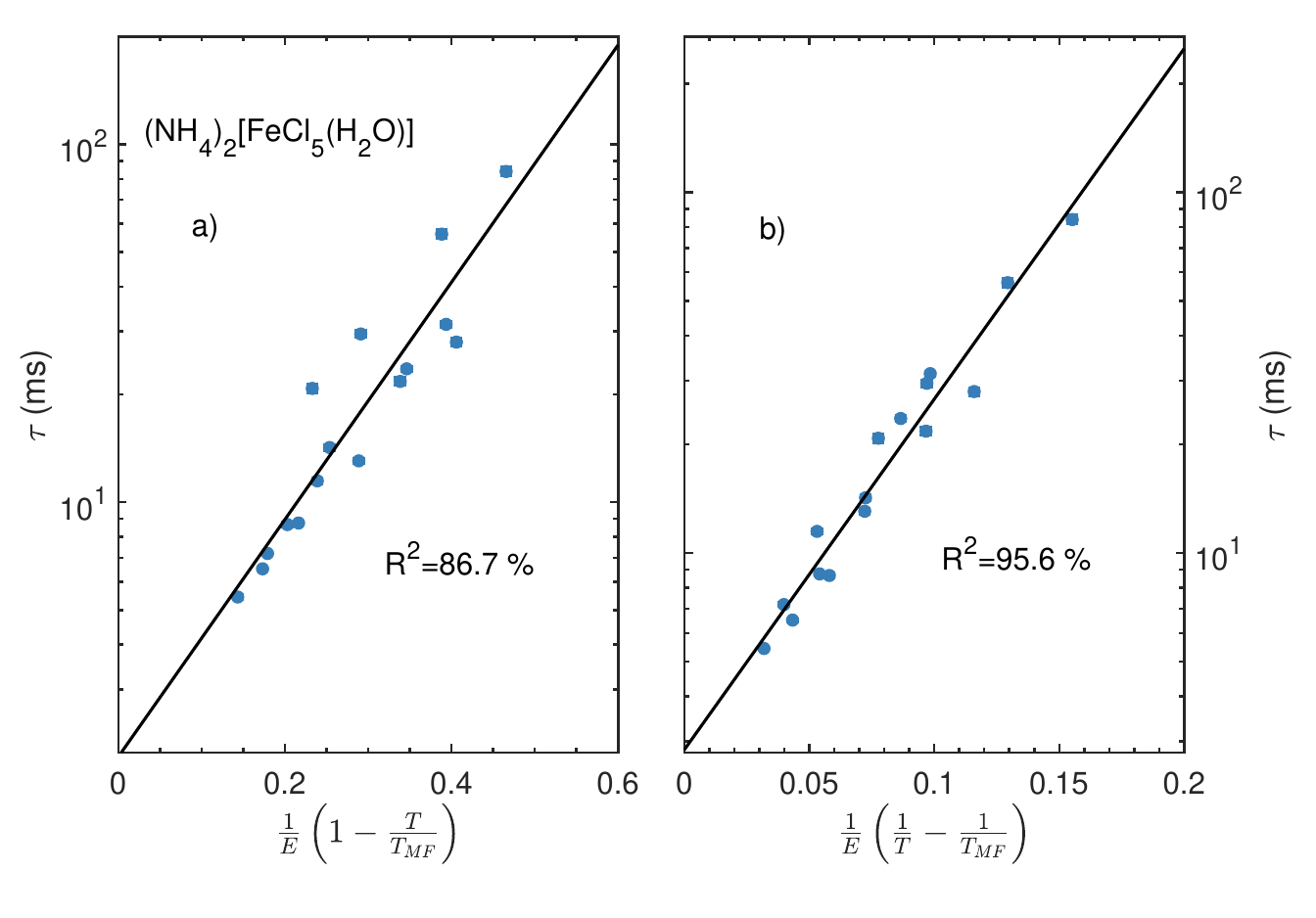}
  \caption{\label{fig:scaling}The different scaling concepts for the temperature and electric-field dependent multiferroic domain inversion are compared. 
  	Panel a) displays the relaxation times plotted against the argument  $\frac{1}{E}(1-\frac{T}{T_{MF}})=\frac{T_r}{E}$ that is observed in some ferroelectric materials, and  panel b) shows the scaling with the argument  $\frac{1}{E}(\frac{1}{T}-\frac{1}{T_{MF}})=\frac{T_r}{ET}$ predicted by the Arrhenius-Merz combination, see equation \ref{eq:arrhenius-merz}. Quantitative agreement factors are included.}
  \label{fig:scaling}
 \end{figure}

All fitted relaxation times are shown in Fig. \ref{fig:relaxation_times} as a function of inverse temperature and inverse electric field. 
For the sake of simplicity, only $\tau_b$ is plotted as both relaxation times $\tau_a$ and $\tau_b$ differ approximately by a factor of 2-3 
due to the observed asymmetry of domain inversion (see Fig. \ref{fig:switching}). 
However, the following discussion is qualitatively the same for both directions.
The electric-field and temperature dependence of the multiferroic relaxation time in \nh follows again the combined Arrhenius-Merz law (Equation \ref{eq:arrhenius-merz}).
Fitting all determined relaxation times with Equation \ref{eq:arrhenius-merz} yields the values $A_0=\SI{22.4(14)}{\kelvin\kilo\volt\per\milli\meter}$ and $\tau^*=\SI{2.85(31)}{\milli\second}$ for the multiferroic domain relaxation in \nhkomma{.}
The measured relaxation times cover two orders of magnitude in time as a function of temperature and electric field. 
Astonishingly, the fastest resulting relaxation time $\tau^*$ in molecular \nh is one order of magnitude slower than that in \tbmno \cite{stein2020} and more than two orders of magnitude slower than that in \nafegeo \cite{biesenkamp2021b}. 
However, this has to be set in relation to the multiferroic transition temperature and to the activation constant. 
\tbmno exhibits a much higher transition temperature and the two order of magnitude larger activation constant $A_0$ tremendously slows down the relaxation process at low temperature. 
In the multiferroic temperature range of \nhkomma{,} the \tbmno domain inversion is thus much slower. The multiferroic transition temperature of \nafegeo is  higher than that in \nhkomma{,} whereas the activation constant is smaller. Therefore, the relaxation in this oxide system is still faster compared to \nhkomma{,} when considering the same temperature range. The molecular constituents in \nh thus do not imply faster relaxation. 

For \nafegeo  the critical relaxation time $\tau^*$ and thus the fastest possible multiferroic domain inversion agrees with the order of magnitude of the spin wave velocity \cite{biesenkamp2021b}, $v_{sw}\sim1000$\,m/s. The inversion of the multiferroic domains is associated with the inversion 
of the chirality of cycloidal order. Among the several low-energy magnons in such a cycloidal arrangement, the phason modes and their spin-wave velocity seem most related to such an effect \cite{Senff2007,Holbein2023}. 
Also for \nd the magnon dispersion was studied by inelastic neutron experiments revealing spin-wave velocities of the same
order of magnitude in the multiferroic phase \cite{Bai2021}. However, it is important to emphasize that in an antiferromagnetic material with a spin gap, spin-wave velocities can be deduced only away from the incommensurate magnetic zone centers. 
Right at the zone center, the magnon dispersion becomes flat and the derivative $\frac{dE}{dq}$ depends on the
propagation vector. 
Therefore, the effective spin-wave velocity being relevant for the multiferroic domain relaxation can
be much slower than that observed in the magnetic Brillouin zone. 
A detailed determination of the spin gap associated with the phason branch was only reported for multiferroic TbMnO$_3$ \cite{Senff2007,Holbein2023}. Although the phason excitations extend to very low
energies of the order of 0.1\,meV, the phason right at the zone center exhibits a clear tail to larger
energies that can arise from the interplay between Mn and Tb moments. A larger gap of the phason branch in \nh or in any other multiferroic material can induce a much flatter magnon dispersion and thus a much smaller effective spin-wave velocity that slows the relaxation down. Further high-resolution inelastic neutron scattering experiments on \nh and other type-II multiferroic materials are highly desirable.
The existing data on TbMnO$_3$ \cite{stein2020}, \nafegeo \cite{biesenkamp2021b} and  \nh already 
exclude a simple relation between $\tau^*$ and the spin-wave velocities at larger propagation vector.

The validity of the combined Arrhenius-Merz relation is confirmed by the implied scaling of the relaxation time $\tau=f(T_\text{r}/(ET))$, which is displayed in Fig. \ref{fig:scaling} b). Scaling against the argument of the Arrhenius-Merz relation is well fulfilled down to low temperature. In contrast, the relation
$\tau=f(T_\text{r}/E)$ proposed for proper ferroelectrics \cite{scott2000} yields stronger deviations, see  Fig. \ref{fig:scaling} a). There is no indication for a failure of the relation at the lowest studied temperatures. In particular relaxation times are not getting shorter as one expects for either depinning of domain walls or for the occurrence of
quantum mechanical tunneling phenomena \cite{Kagawa_2016,Brooke_2001}.

\section{Higher harmonics}

\begin{figure}
 \includegraphics[width=\columnwidth]{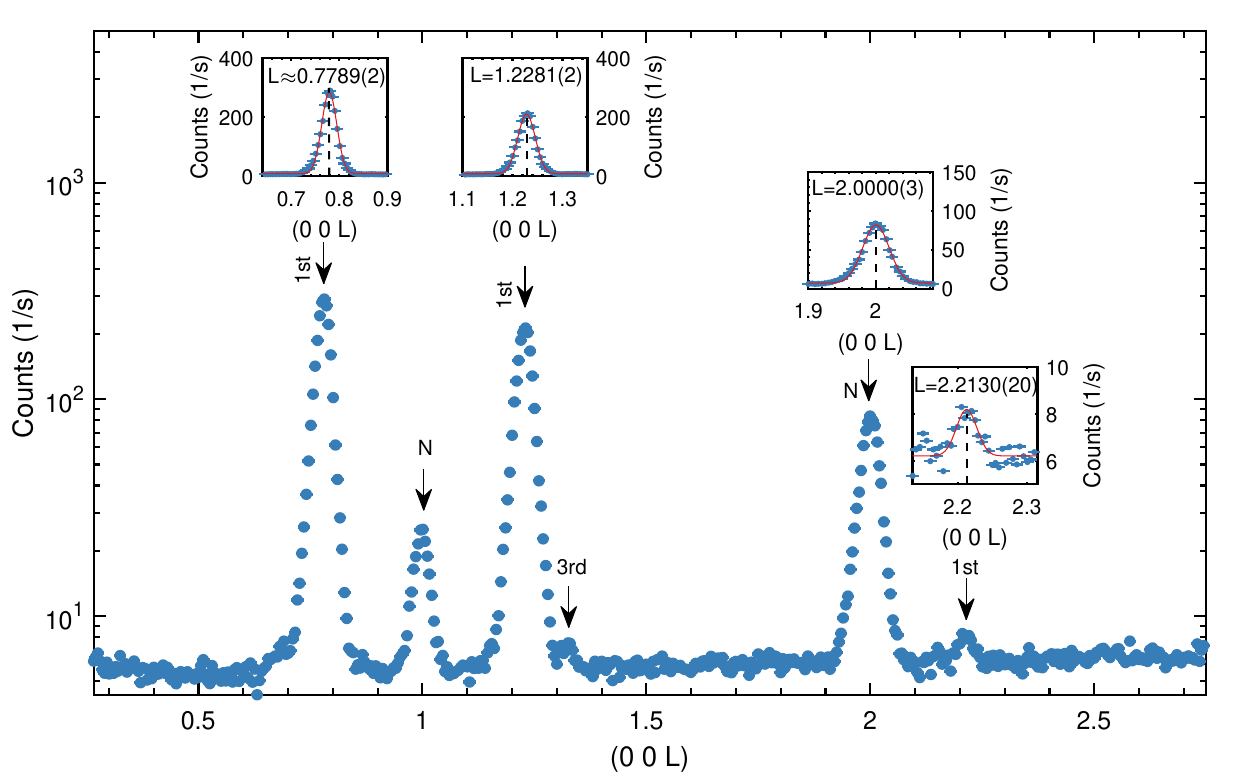}
  \caption{\label{fig:L-scan} $Q_L$-scan across (0 0 L) direction at T = 1.48K. The
observed nuclear (N), magnetic first order (1st) and magnetic third order (3rd)
reflections are marked by arrows and the shown insets present Gaussian fits
for some of the incommensurate first order and nuclear reflections, whereby
the value of the fitted peak position is respectively included in all panels. An
equivalent scan for the deuterated compound at $T$ = 1.5\,K can be found in Ref. \cite{Tian2016}. }
  \label{00l}
 \end{figure}

\begin{figure}
 \includegraphics[width=\columnwidth]{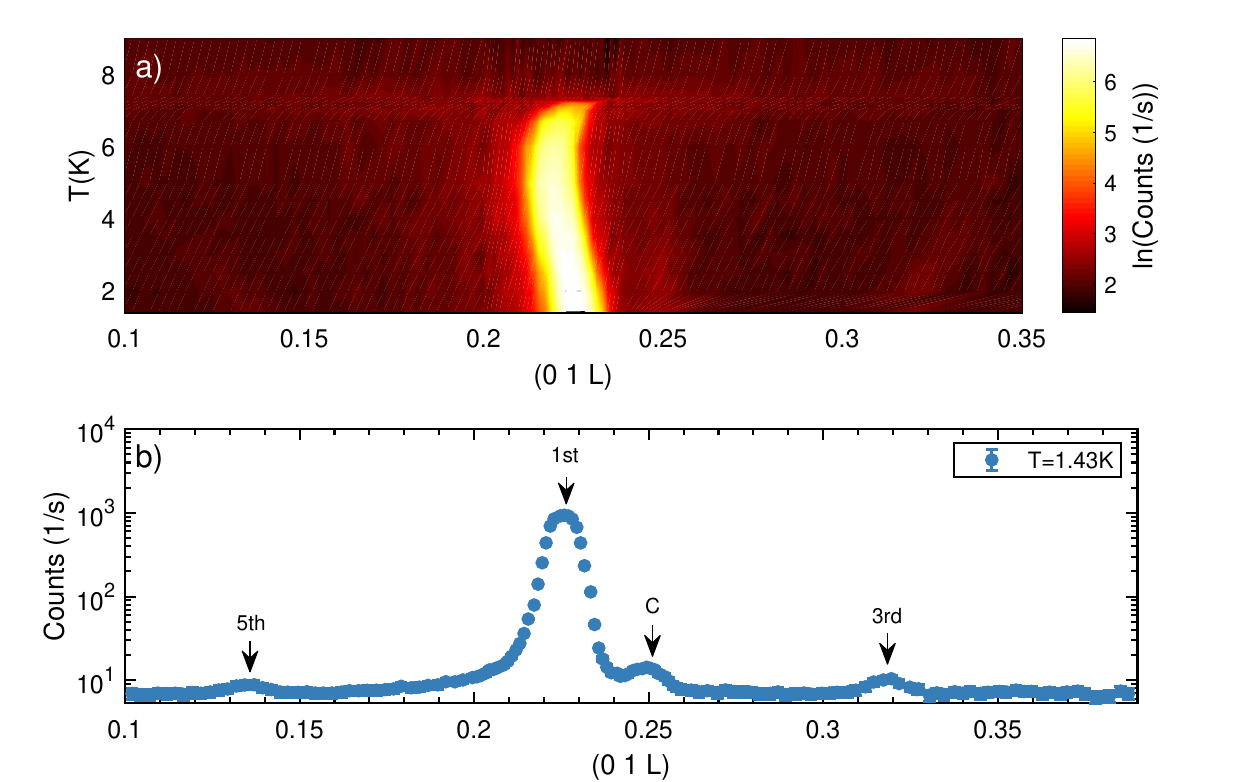}
  \caption{\label{fig:harmonics}The upper panel a) shows a temperature dependent $Q_L$ scan across the magnetic reflection $\vec{Q}=(0\:1\:0.23)$.
  One single scan is shown in panel b) for $T$\,=\,1.43\,K, where first (1st), third (3rd), fifth (5th) order harmonics and commensurate (C)
  contributions are marked by arrows. }
  \label{01l}
 \end{figure}

\begin{figure}
 \includegraphics[width=\columnwidth]{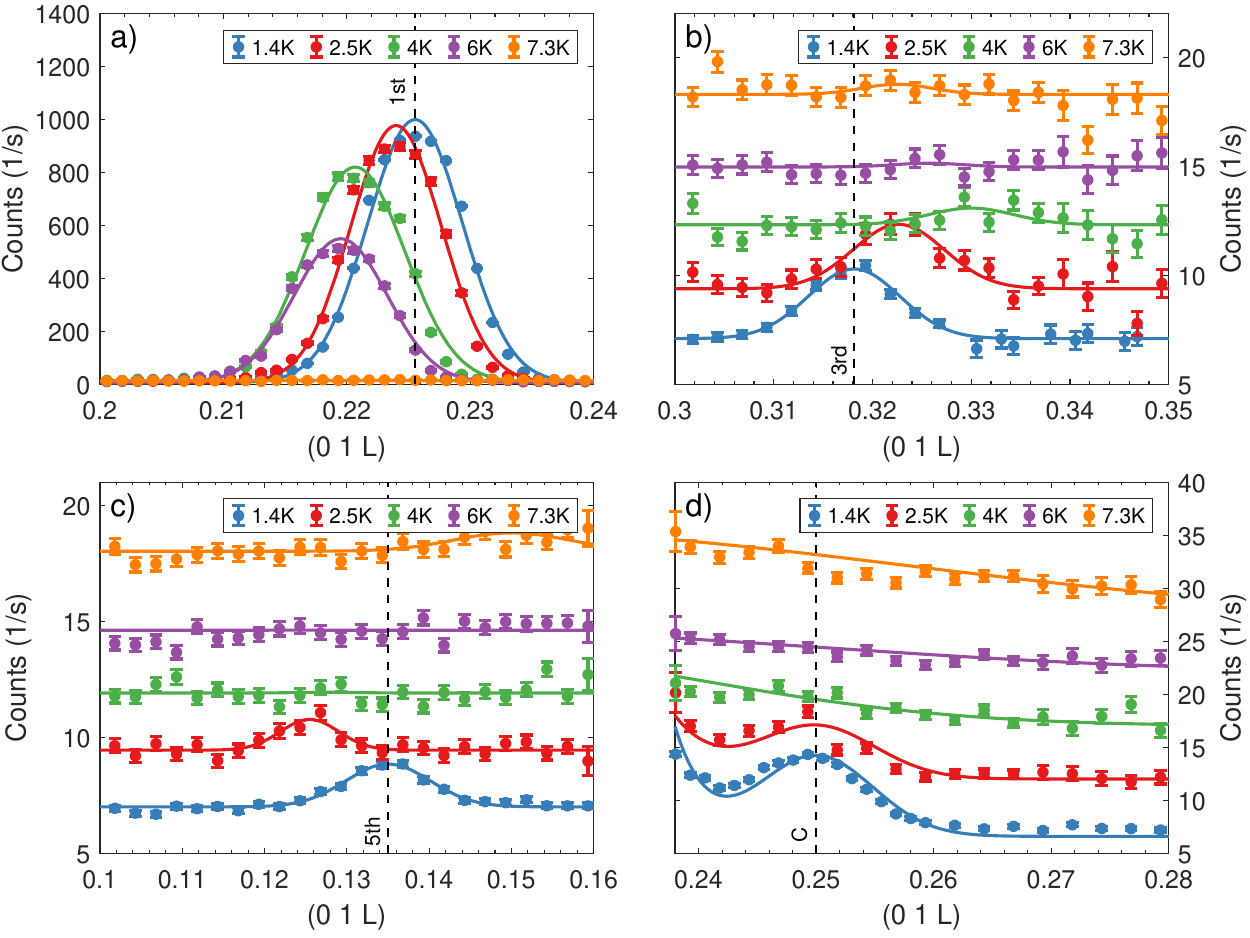}
  \caption{\label{fig:profiles}The observable reflections from $L$ scans and the respective
fits. The first, third and fifth order reflections are respectively plotted in a)-c)
together with a Gaussian fit, for which a constant background was assumed.
This contrasts the fit of the commensurate reflection, which is displayed in
d). Here, the background was assumed to be the tail of a second Gaussian
function and the commensurate peak position was fixed to L = 0.25. For a
better view, all respective scans in b)-d) are shifted by a constant offset with
respect to the y-axis. }
  \label{zoom01l}
 \end{figure}

\begin{figure}
 \includegraphics[width=\columnwidth]{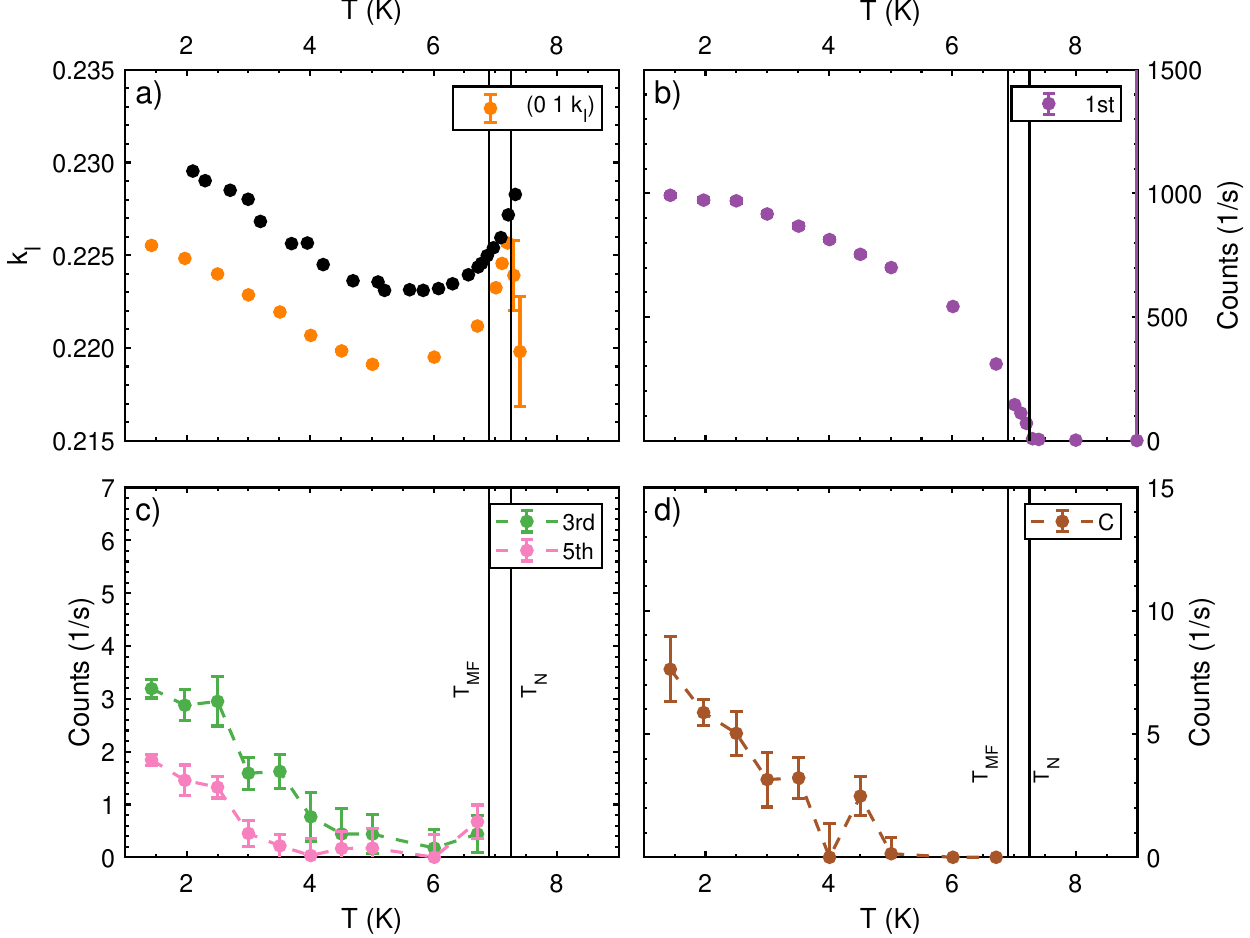}
  \caption{\label{fig:fitresults}Results from temperature dependent fits
of all harmonic and commensurate reflections. Panel a) displays the obtained
temperature dependence of the measured incommensurate propagation vector
(orange data points) and compares it with data (black data points) from
Ref. \cite{Tian2016} obtained on a deuterated sample. The plots in
b)-d) show the temperature dependent intensity of the 1st, 3rd, 5th harmonics
as well as that of the commensurate (C) reflection for \nhkomma{.} }
  \label{fitresults}
 \end{figure}

Tian \textit{et al.} reported the competition of various ordering schemes \nd : incommensurate modulations with even and odd harmonics
as well as a commensurate ordering appearing at low temperature ($T=\SI{1.5}{\kelvin}$) \cite{Tian2016,Tian2018}.
The interference of incommensurate and commensurate ordering can depin multiferroic domain walls and was
suggested to cause the anomalous relaxation behavior in \mnwo \cite{biesenkamp2020,Baum2014,Niermann_2014,Hoffmann2011a}.
Tian et al. utilized neutron polarization analysis to separate the respective magnetic and nuclear contributions. 
Odd order harmonics and the commensurate signal are solely of magnetic origin, whereas the second order reflection is mainly of nuclear origin but possesses also a finite magnetic contribution \cite{Tian2016}. 
This closely resembles the occurrence and origin of even and odd harmonics in \mnwo \cite{Finger2010,biesenkamp2020},
in particular the double, nuclear and magnetic, character of the second order modulation.
The appearance of second-order harmonics is rather common, because every incommensurate magnetic modulation implies a nuclear lattice modulation of half the period in real space due to exchange striction. 
Higher order contributions were also studied in other multiferroics \cite{Kajimoto2004,Kenzelmann2005,Wilkins2009} as well as in Ba$_3$NbFe$_3$Si$_2$O$_{14}$ that is not multiferroic but exhibits a monodomain chiral magnetic structure inscribed by its non-centrosymmetric crystal structure \cite{Scagnoli2013}. In all these materials, the 3rd order contributions show  peculiar temperature dependencies with a positive curvature that do not follow the temperature dependence of the first-order intensities. 
Moreover, in \nd and similarly in \mnwo \cite{Finger2010,biesenkamp2020} the reported temperature dependence revealed a strong enhancement of second order intensity with the onset of ferroelectric and chiral magnetic ordering, which indicates significant magnetoelastic coupling in the multiferroic phase. The observed third and fifth order harmonics signal a squaring up of the spin spiral. This was also observed in \mnwo as a precursor of the low-temperature commensurate spin $uudd$ phase. 
In \ndkomma{,} no commensurate low-temperature phase was so far reported but even harmonics and a sizable peak intensity for the commensurate reflection $\vec{Q}=(0\:1\:0.25)$ in the incommensurate multiferroic phase indicate that commensurate order coexists and competes with the incommensurate ordering \cite{Tian2016}.
The incommensurate ordering is dominant in zero field but the coexisting commensurate ordering can be stabilized with increasing magnetic field, which leads to an intermediate region between $\SI{1.5}{\tesla}$-$\SI{2.5}{\tesla}$ before the spin spiral becomes commensurate and distorted at $\SI{2.5}{\tesla}$ \cite{Tian2016,Tian2018,Velamazan2018}. 
So far only the temperature dependence of even harmonics in the deuterated material was reported \cite{Tian2016,Tian2018}, but to assess the impact of interfering commensurate and incommensurate ordering on the multiferroic relaxation behavior, temperature dependent measurements of odd harmonics and of the commensurate contributions are required.

% $k_l$
% incommensurate propagation vector $k_{\text{inc}}=(0\:0\:k_l)$  \cite{Tian2016} in Fig \ref{fig:harmonics}
%  $T_\text{S}=\SI{79}{\kelvin}$ \cite{bruening2020}
%  \nh  \nhkomma{,}

Figure \ref{fig:L-scan} presents a $Q_L$-scan across $Q$ = (0 0 L) in \nhkomma{,} which was measured at the lowest
temperature T = 1.48\,K and which resembles the respective measurement from Tian et
al. \cite{Tian2016} on the deuterated sample. 
As already suggested by the lack of isotope effect for the magnetic transition temperatures \cite{bruening2020}, this suggests a minor role of the isotope for the magnetic ordering. 
Nevertheless, some finite impact of the isotope on the magnetic structure is observed and will be
described below. 
In both compounds, nuclear as well as magnetic first- and third-order
reflections exhibit finite intensity at this temperature. The three
incommensurate first order reflections in Fig. \ref{fig:L-scan} are fitted by single Gaussian functions to determine their exact position and thus the incommensurability of the underlying magnetic
structure. The weighted average of determined values for the incommensurability yields the
propagation vector  $k_{\text{inc}}$ = (0 0 0.2245(5)), which differs by approximately 2.5\% \ from the
one, which was reported for the deuterated sample \cite{Tian2016}. This difference cannot be related
to a potential misalignment of the sample as the observed nuclear reflections (0 0 2) is
observed exactly at the expected position $L$ = 2.0000(3) (see Gaussian fit in the right inset
of Fig. \ref{fig:L-scan}). 
%This isotope dependent incommensurability will be further discussed in the
%context of temperature dependent measurements of first-order, third-order, fifth-order and
%commensurate reflections, for which it was seen promising to record Q-scans across (0 1 L).
%This scan direction was chosen as the reported equivalent scan at 1.5K on the deuterated compound \cite{Tian2016} revealed finite
%and significant intensities for the commensurate and for the odd-harmonic reflections.

Temperature-dependent scans are shown in Fig. \ref{fig:harmonics} a), whereby
Fig. \ref{fig:harmonics} b) displays an exemplary single scan at T = 1.5K. This scan clearly confirms
the reported presence \cite{Tian2016} of odd harmonics and also the coexistence of incommensurate and
commensurate reflections in the hydrogen system. 
We find a significantly smaller commensurate contribution that can be quantified by the ratio of (0 1 0.225) and (0 1 0.25)
peaks, which amounts to $\sim$5\% in the deuterated compound \cite{Tian2016} and to less than one percent in the hydrogen crystal.
The coexistence of the commensurate phase seems thus to be sample dependent, and most likely the enhanced disorder
of the deuterated sample stabilizes a larger commensurate contribution.

All peaks were fitted separately by Gaussian functions, whereby a constant background was
assumed for all higher harmonics. In contrast, for the commensurate reflection a second
Gaussian function was used to describe the background, respectively the superimposed
signal from the first order signal. These fits are shown in Fig.  \ref{fig:profiles}. The peak center position
gives the value $k_l$ for the incommensurate propagation vector $k_{\text{inc}}$\,=\,(00$k_l$). Its temperature
dependence is compared with the reported values from a deuterated sample \cite{Tian2016} in Fig.
 \ref{fig:fitresults} a) and obviously, both values for the incommensurability along $c$ differ approximately
by 2\% in the entire temperature range. An isotope effect was so far
reported only for the structural order-disorder transition temperature
$T_S$ = 79\,K, while isotope exchange does not significantly change the magnetic transition temperatures \cite{bruening2020}. 

For assessing the intensity of higher harmonic reflections and hence the magnitude of
anharmonic contributions to the spiral ordering, the related first order intensity is plotted
in Fig. \ref{fig:fitresults} b) as a function of temperature. 
In Fig. \ref{fig:fitresults} c) the temperature dependencies of
both odd higher harmonics are shown. 
The third order reflection evolves below $T$ $\sim$ 6\,K, whereas the fifth order reflection develops only below $T$ $\sim$ 4\,K. 
These intensities are of the same order of magnitude and they saturate at low temperature. 
By considering the geometry factor, it turns out that the peak intensity at $T$ = 1.5\,K amounts to approximately $10^{-3}$ of the
incommensurate first order reflection, wherefore these odd harmonics can be estimated to
be even relatively stronger than those in MnWO$_4$ \cite{biesenkamp2020}. 
The related squaring up of spins enforces
tiny commensurate fragments which is further indicated by the simultaneous development
of a commensurate reflection (see Fig. \ref{fig:fitresults})).
However, the interfering commensurate
ordering does not affect the relaxation behavior above $T$ = 3\,K (see Fig. 4). 

The main difference between \nh and MnWO$_4$ \cite{biesenkamp2020} is the fact that
odd harmonics and the development of tiny commensurate fragments appear at much
lower temperature in the molecular compound. The pinning strength at lower temperature is naturally stronger and is
thus capable to compensate the depinning effects from interfering commensurate ordering.
Further, the intensity of odd harmonics in MnWO$_4$ diverges at the lower transition to
commensurate order \cite{biesenkamp2020}, whereas in \nh the respective intensity follows
the intensity of the incommensurate first order peak and saturates at low temperature.
This behavior agrees with the absence of a transition to pure commensurate ordering at
low temperature in \nhkomma{.} It is thus expected that depinning effects do not significantly grow
at much lower temperatures in \nhkomma{,} which explains that the relaxation
behavior agrees with the combined Arrhenius-Merz law even in the
presence of some interfering commensurate ordering at 3\,K. 

TbMnO$_3$ is another multiferroic material, whose domain relaxation was studied \cite{stein2020}, and it
also exhibits the appearance of higher order contributions \cite{Kajimoto2004,Kenzelmann2005,Wilkins2009}, but the use of rather large crystals in the neutron
experiments prohibits a quantitative analysis, see the discussion in Ref. \cite{Holbein2023}.

 \section{Conclusion}

The multiferroic domain relaxation and the temperature dependence of magnetic correlations were studied in the molecular system \nh on a hydrogen-containing crystal, in contrast to most reported neutron diffraction experiments on this system analyzing deuterated crystals
\cite{Velamazan2015,Tian2016,Tian2018,Velamazan2017,Velamazan2018,bruening2020}.
The analysis of the hydrogen-containing sample reveals a slight isotope effect on the incommensurability of the magnetic structure. 
In addition the commensurate ordering coexisting at low temperature with the main incommensurate structure 
is considerably suppressed compared to measurements on deuterated samples. 
As deuteration intrinsically induces some disorder both effects can essentially stem from these perturbations.

Over a significant field and temperature range, the temperature and electric-field dependence of multiferroic relaxation in \nh
follows the simple combined Arrhenius-Merz relation. The domain kinetics in the
molecular compound are thus similar to the relaxation behavior in the previously studied oxide systems
TbMnO$_3$ and NaFeGe$_2$O$_6$ \cite{stein2020,biesenkamp2021b}, but astonishingly, the relaxation in the molecular system is
much slower than in NaFeGe$_2$O$_6$ in spite of similar transition temperatures and similar spin-wave velocities indicating that other quantities, such as the spin gap, also play an important role.

\section{\label{sec:level1}Acknowledgements}

% acknowledgement!!!!!!!!!!!!!!!!!!!!!!!!!!!!!!!!!!!!!!!!!!!!!!!!!!!!!!!!!!!!!!!!!!!!

This work was funded by the Deutsche Forschungsgemeinschaft (DFG,
German Research Foundation) - Project number 277146847 - CRC 1238, projects A02 and B04.
The neutron scattering data from the IN12 and IN3 diffractometer are available \cite{dataIN12_CRG2573}.

%\nocite{apsrev41Control}
%\bibliographystyle{apsrev4-1}

%\bibliographystyle{plain}
%\bibliographystyle{apsrev4-2}

%\bibliography{revtex-control,NH4FeCl5H2O}

%\begin{thebibliography}{46}%

%\end{thebibliography}%
%

\end{document}